\documentclass[onecolumn]{revtex4}

\def\al{\alpha}
\def\be{\beta}
\def\ga{\gamma}
\def\de{\delta}
\def\ep{\epsilon}

\def\et{\eta}

\def\la{\lambda}

\def\ph{\phi}

\def\ps{\psi}

\def\fr#1#2{{{#1} \over {#2}}}

\def\frac#1#2{{\textstyle{{#1}\over {#2}}}}

\def\vev#1{\langle {#1}\rangle}

\def\lsim{\mathrel{\rlap{\lower4pt\hbox{\hskip1pt$\sim$}}
    \raise1pt\hbox{$<$}}}
\def\gsim{\mathrel{\rlap{\lower4pt\hbox{\hskip1pt$\sim$}}
    \raise1pt\hbox{$>$}}}
\def\sqr#1#2{{\vcenter{\vbox{\hrule height.#2pt
         \hbox{\vrule width.#2pt height#1pt \kern#1pt
         \vrule width.#2pt}
         \hrule height.#2pt}}}}

\def\pt#1{\phantom{#1}}

\def\vb#1#2{e_{#1}^{{\pt{#1}}#2}}

\def\lvb#1#2{e_{#1#2}}

\def\barvb#1#2{\bar e_{#1}^{{\pt{#1}}#2}}

\def\barlvb#1#2{\bar e_{#1#2}}

\newcommand{\beq}{\begin{equation}}
\newcommand{\eeq}{\end{equation}}
\newcommand{\bea}{\begin{eqnarray}}
\newcommand{\eea}{\end{eqnarray}}
\newcommand{\bit}{\begin{itemize}}
\newcommand{\eit}{\end{itemize}}
\newcommand{\rf}[1]{(\ref{#1})}

\begin{document}

\title{Gravity Theories with Background Fields and \\
Spacetime Symmetry Breaking}

\author{Robert Bluhm}

\affiliation{
Physics Department, Colby College,
Waterville, ME 04901 
}

\begin{abstract}
An overview is given of effective gravitational field theories
with fixed background fields that break spacetime symmetry.
The behavior of the background fields and the types of excitations
that can occur depend on whether the symmetry breaking
is explicit or spontaneous.  
For example, when the breaking is spontaneous,
the background field is dynamical and
massless Nambu--Goldstone and massive Higgs excitations can appear.
However,
if the breaking is explicit, 
the background is nondynamical,
and in this case additional metric or vierbein excitations occur 
due to the loss of local symmetry,
or these excitations can be replaced by dynamical 
scalar fields using a St\"uckelberg approach.
The interpretation of Noether identities that must hold in each case differs,
depending on the type of symmetry breaking,
and this affects the nature of the consistency conditions that must hold. 
The Noether identities also shed light on why the St\"uckelberg approach works,
and how it is able to restore the broken spacetime symmetry in a theory with explicit breaking.
\end{abstract}

\maketitle

\section{Introduction}

Some of the most important open questions in physics concern gravity.
These include the question of how General Relativity (GR) merges with the
Standard Model (SM) of particle interactions in a quantum theory.
Additional open questions concern the nature of dark energy and dark matter
in gravity theories and cosmology.
In many cases, investigations of these questions involve looking at modified
gravitational and particle interactions,
many of which are described by effective field theories that include fixed background fields.
These background fields break spacetime symmetries,
such as local Lorentz invariance and diffeomorphism invariance.
These breakings can occur spontaneously,
where dynamical tensor fields acquire a nonzero vacuum value,
or explicitly, when nondynamical background tensors are included directly
in the Lagrangian \cite{rb15a}.

Examples of effective theories involving background fields
in the context of gravity include
the Standard-Model Extension (SME) \cite{sme1,sme2,sme3,akgrav04,rbsme}, 
Bumblebee models~\cite{ks1,ks2,ks3,bb1,bb2,rbngrpav,ms1,ms2,ch14,eu15} or Einstein--Aether models~\cite{ee1,ee2}, 
Cardinal models \cite{cardinal}, models with an antisymmetric two-tensor \cite{phon1,phon2}, 
Chern--Simons gravity \cite{rjsp1,rjsp2}, massive gravity \cite{mg1,mg2},
and theories with spacetime-varying couplings \cite{varcoups1,varcoups2,tvar1,tvar2,tvar3}.
In some of these examples, the spacetime symmetry breaking is spontaneous,
while in others it is explicit.
In certain cases, either type of breaking or even a combination of 
both types of symmetry breaking can occur \cite{rb15b}.

Despite the presence of background fields that break spacetime symmetries,
a meaningful physical theory must still be observer independent \cite{sme1,sme2}.
This means that the choice of spacetime coordinates or local Lorentz bases cannot
influence the underlying physics.
In the context of a gravity theory,
this requires that general coordinate invariance and the passive form of 
local Lorentz invariance must still~hold.

A useful distinction can therefore be made between what are called particle
and observer spacetime transformations.
Particle transformations act (in an active sense) on physical fields but not background fields,
while observer transformations act (in a passive sense) on all fields including the background fields.
It is the particle symmetries consisting of diffeomorphism invariance and local Lorentz invariance
that are broken either spontaneously or explicitly by the presence of background fields,
while the mathematical observer symmetries must continue to hold.

If the symmetry breaking is spontaneous,
the action describing the theory remains invariant under both the particle and observer
transformations when all of the Nambu--Goldstone (NG) and massive Higgs-like excitations are included.
However,
with explicit breaking,
the particle symmetries do not hold.
Nonetheless, the observer symmetries must remain mathematical symmetries of the action
in order to maintain observer independence.
This then sets up a potential conflict between the broken particle symmetries
and the unbroken observer symmetries when the symmetry breaking is explicit.
However, theories with spontaneous spacetime symmetry breaking do not 
encounter such conflicts \cite{akgrav04}.

In the context of a gravitational theory,
which must respect geometrical identities such as the Bianchi identities,
the conflicts that arise with explicit breaking
can lead to theoretical inconsistency unless certain conditions hold.
One approach for obtaining useful consistency conditions  
is to look at the mathematical Noether identities
associated with the observer invariances \cite{rbas16}.
Using these identities,
the question of whether a particular theory is consistent 
or not can then be examined.

In this overview,
a general treatment is used to examine different features and behaviors that can arise
in gravity theories that contain background fields,
including their dependence on whether the spacetime symmetry breaking is spontaneous or explicit.
First, in the next section,
the properties of the background fields are examined for
the two types of symmetry breaking.
Next, in Section \ref{sec3},
the different types of excitations that can occur in
conjunction with the symmetry breaking are investigated and discussed.
These include massless Nambu--Goldstone (NG) and massive Higgs-like
modes in the case of spontaneous breaking,
or additional metric modes or St\"uckelberg fields in
the case of explicit breaking.
Section \ref{sec4} looks at the Noether identities that hold 
and how their interpretation depends on the form of the symmetry breaking.
A summary and conclusions are given in Section \ref{sec5}.

It is important to keep in mind that a major component of the research effort devoted to
testing spacetime symmetries consists of experimental tests of local Lorentz symmetry and GR
\cite{CPTmeetings,m05,ML12,JT14,RB14,Iorio1,will, hees,smoot,stubbs,mueller1,mueller2,iorio2,yagi,shao,angonin1,angonin2}.
The~SME is widely used as the phenomenological framework for these tests,
and the sensitivities to Lorentz violation are expressed 
as experimental bounds on the SME coefficients
\cite{aknr-tables}.
This overview will not discuss experimental tests of Lorentz symmetry.
However,
the question of how background fields are interpreted depending
on the type of symmetry breaking is relevant to the SME.
For example, in the SME restricted to Minkowski spacetime,
the coefficients for Lorentz violation can be treated as fixed 
nondynamical background fields that explicitly break Lorentz symmetry.
However,
in the gravity sector of the SME 
\cite{qbak06,akjt1,akjt2,akjt3,hb14,qbakrx,cl16,combined,ksmm1,ksmm2,ksmm3}
more caution is usually used in the case of explicit breaking.
Typically, the pure-gravity sector of the SME assumes the background SME coefficients
are dynamical vacuum solutions
and that the NG and Higgs modes must be accounted for in 
order to avoid potential inconsistency issues.

\section{Background Fields} \label{sec2}

A variety of background fields are considered in effective gravitational field theories.
These include fixed nondynamical background scalar or tensor fields,
which explicitly break diffeomorphism invariance and local Lorentz invariance.
Alternatively,
dynamical background fields can arise as vacuum expectation values
if the symmetry breaking is spontaneous.

To consider these types of models in a general way,
and to examine the differences between explicit and spontaneous 
spacetime symmetry breaking,
let $\bar k_{\la\mu\nu\cdots}$ denote a generic background field.
Consider it as a fixed scalar or tensor with an unspecified number of components.
An effective gravitational theory containing such a background also includes interactions
with other fields that are fully dynamical, 
including the metric, $g_{\mu\nu}$, and conventional matter fields.
The latter are denoted generically as $f^\ps$, 
where $\ps$ is a collective label for the all of the matter-field tensor indices.
Assuming an Einstein--Hilbert form for the pure-gravity sector,
and using units with $8 \pi G = 1$,
the Lagrangian for a theory of this form can be written as
\beq
S= \int d^4x \sqrt{-g} \left[ \fr 1 2 R + {\cal L}(g_{\mu\nu}, f^\ps, \bar k_{\la\mu\nu\cdots}) \right] .
\label{brokenSmetric}
\eeq
The equations of motion are obtained by varying $S$ with respect to the dynamical fields.
This yields the Einstein equations, $G^{\mu\nu} = T^{\mu\nu}$,
when variations with respect to $g_{\mu\nu}$ are performed.
Variations with respect to the matter fields $f^\ps$ yield the Euler--Lagrange equations for $f^\ps$,
which can be denoted generically as $\fr {\de {\cal L}} {\de f^\ps} = 0$.
These equations typically include partial derivative contributions.
For example, if ${\cal L}$ depends on both $f^\ps$ and $D_\mu f^\ps$,
then the Euler--Lagrange expression is
\beq
\fr {\de {\cal L}} {\de f^\ps} \equiv - D_\mu \left( \fr {\partial {\cal L}} {\partial D_\mu f^\ps} \right) + \fr {\partial {\cal L}} {\partial f^\ps} ,
\label{EL}
\eeq
where $D_\mu$ is a covariant derivative.

The question of whether the background field obeys Euler--Lagrange equations or not depends
on the type of symmetry breaking.
With explicit breaking,
the background is nondynamical,
and therefore Euler--Lagrange equations need not hold.
Thus, in general
\beq
\fr {\de {\cal L}} {\de \bar k_{\la\mu\nu\cdots}} \ne 0 .
\label{nondyn}
\eeq
However, if the symmetry breaking is spontaneous,
then the background $\bar k_{\la\mu\nu\cdots}$ arises as
a vacuum solution for a dynamical field $k_{\la\mu\nu\cdots}$,
which allows it to be written as
$\bar k_{\la\mu\nu\cdots} = \vev{\bar k_{\la\mu\nu\cdots}}$.
In this case,
$\bar k_{\la\mu\nu\cdots}$ is a solution of the vacuum Euler--Lagrange equations,
\beq
\left( \fr {\de {\cal L}} {\de \bar k_{\la\mu\nu\cdots}} \right)_{\rm vacuum} = 0 .
\label{dynvac}
\eeq
Away from the vacuum solution,
the dynamical field $k_{\la\mu\nu\cdots}$ has excitations in 
the form of massless NG and massive Higgs-like modes.
When these are included in the effective theory,
particle diffeomorphism and local Lorentz invariance still hold,
and $k_{\la\mu\nu\cdots}$ is a solution of its Euler--Lagrange~equations.

\subsection{Diffeomorphism Breaking}

Since the metric and conventional matter fields are fully dynamical,
they transform under particle diffeomorphisms in the usual way,
with infinitesimal changes given by Lie derivatives defined
with respect to a spacetime vector $\xi^\mu$.
For example,
the metric transforms as 
\beq
g_{\mu\nu} \rightarrow g_{\mu\nu} + {\cal L}_{\xi} g_{\mu\nu}
= g_{\mu\nu} + D_\mu \xi_\nu + D_\nu \xi_\mu ,
\label{metricdiff}
\eeq
while the matter fields transform as
\beq
f^\ps \rightarrow f^\ps + {\cal L}_{\xi} f^\ps  .
\label{fdiff}
\eeq
However,
the background field breaks particle diffeomorphisms and remains fixed
under these transformations, obeying 
\beq
\bar k_{\la\mu\nu\cdots} \stackrel {\rm particle} { \longrightarrow} \, \bar k_{\la\mu\nu\cdots} .
\label{kbarparticletrans}
\eeq

At the same time,
to be physically viable, the theory must be observer independent.
This requires that the action remains invariant under general coordinate transformations.
For infinitesimal coordinate transformations,
$x^\mu \rightarrow x^\mu - \xi^\mu$,
defined using vectors, $-\xi^\mu$, in the inverse direction,
all of the tensor fields in the theory transform mathematically 
with changes given by Lie derivatives.
For this reason, these transformations are referred to as observer diffeomorphisms.
The metric and conventional matter fields transform the same way under these infinitesimal
coordinate transformations as they do under particle diffeomorphisms.
Note, however, that under these observer transformations, 
the background field also transforms,
obeying
\beq
\bar k_{\la\mu\nu\cdots} \stackrel {\rm observer} { \longrightarrow} \, \bar k_{\la\mu\nu\cdots} + {\cal L}_{\xi} \bar k_{\la\mu\nu\cdots} .
\label{kbarobsvtrans}
\eeq

If the diffeomorphism breaking is explicit,
then the action is not invariant,
and $\de S \ne 0$ under particle diffeomorphisms.
Nonetheless, the action is still required to obey $\de S = 0$ mathematically
under the observer diffeomorphisms in order to maintain observer independence.
It is in this way that the explicit breaking of diffeomorphisms while maintaining observer
independence can lead to a potential conflict.
To be fully consistent, a theory must resolve or evade this conflict.

With explicit diffeomorphism breaking,
four local gauge invariances associated with the local vectors $\xi^\mu$ do not occur.
As a result, there are up to four additional degrees of freedom in the metric compared to GR,
which can therefore modify gravitational interactions.
These extra metric modes can also give rise to ghosts,
which is therefore an important consideration in theories with explicit diffeomorphism breaking.

In many theories with explicit diffeomorphism breaking, 
a St\"uckelberg approach is used \cite{ags03}.
In this approach, the background $\bar k_{\la\mu\nu\cdots}$ is rewritten in terms of
four dynamical scalars, $\ph^A$, labeled with an index $A = 0,1,2,3$.
The replacement is given as
\beq
\bar k_{\la\mu\nu \cdots} (x) = \partial_\la \ph^A \partial_\mu \ph^B 
\partial_\nu \ph^C \cdots \bar k_{ABC\cdots} (\ph) .
\label{ktransf1}
\eeq
The four scalars transform under particle diffeomorphisms,
and the substitution of \rf{ktransf1} into the action $S$ is sufficient to restore
particle diffeomorphism invariance.
As a result, four degrees of freedom in the metric can again be
treated as gauge degrees of freedom.
However, the theory still has up to four additional degrees of freedom,
in comparison to GR,
due to the added St\"uckelberg fields.

With spontaneous diffeomorphism breaking,
the number of degrees of freedom in the metric is similar to GR.
This is because particle diffeomorphism invariance still holds when the
NG modes are included in the action.
Thus, there are four local gauge degrees of freedom that can
be used to eliminate four degrees of freedom in the metric.

\subsection{Local Lorentz Symmetry Breaking}

To reveal the local Lorentz invariance,
a vierbein formalism can be used.
In this case,
the metric is replaced by a vierbein $\vb a \mu$,
where the defining relation is
\beq
g_{\mu\nu} = \vb \mu a \vb \nu b \et_{ab} .
\label{gee}
\eeq
Here, Greek indices are used for components defined on
the spacetime manifold,
while Latin indices denote components defined with respect
to a local Lorentz frame.

Again, a distinction can be made between particle 
and observer transformations.
Under particle local Lorentz transformations,
which depend on six antisymmetric parameters $\ep_{ab}$,
the vierbein transforms as a local vector,
\beq
\vb \mu a \rightarrow \vb \mu a + \ep^a_{\pt{a}b} \vb \mu b .
\label{LTe}
\eeq
The matter fields $f^\ps$ have components with respect to the local basis,
which can be denoted generically as $f^y$.
These transform as irreducible representations of the local Lorentz group,
which have the form
\beq
 f^y \rightarrow  f^y + \fr 1 2 \ep^{ab} (X_{[ab]})_{\pt{y}x}^{y} f^x .
 \label{LLT}
 \eeq

Since the background field is fixed,
both the components $\bar k_{\la\mu\nu\cdots}$ defined with respect
to the spacetime coordinate frame
and $\bar k_{abc\cdots}$ defined with respect to the local Lorentz frame
remain unchanged under particle spacetime transformations.
Because the frames themselves do not change either under particle transformations,
the background components are therefore connected by a fixed background vierbein,
denoted as $\barvb a \mu$.
The defining relation for the background vierbein is
\beq
\bar k_{\la\mu\nu\cdots} = \barvb \la a \barvb \mu b \barvb \nu c \cdots \bar k_{abc \cdots} ,
\label{ebar}
\eeq
where each quantity in this expression remains fixed under both particle
diffeomorphisms and local Lorentz transformations.

In a vierbein formalism,
the action replacing \rf{brokenSmetric} can then be written as
\beq
S= \int d^4x e \left[ \fr 1 2 R + {\cal L}(\vb \mu a,\barvb \mu a,f^y, \bar k_{abc\cdots}) \right] ,
\label{brokenSvier}
\eeq
where $e$ is the determinant of the vierbein.
The action in this case depends on the dynamical vierbein and conventional matter fields
as well as the background $\bar k_{abc\cdots}$ and fixed vierbein $\barvb \mu a$.

If the symmetry breaking is explicit,
the background field and the background vierbein are nondynamical,
and the action is not invariant under particle diffeomorphisms and local Lorentz transformations.
However, in order to maintain observer independence,
the action must be mathematically invariant under observer local Lorentz transformations,
consisting of changes of the local Lorentz bases,
as well as observer diffeomorphisms.
In this case, all of the fields,
including $\bar k_{abc\cdots}$ and $\barvb a \mu$, 
transform so as to keep $S$ unchanged.
The combination of broken particle local Lorentz invariance with
unbroken observer local Lorentz symmetry can lead to potential conflicts 
similar to those that occur with explicit diffeomorphism breaking.
Theoretical consistency requires that these conflicts must be avoided as well.  

However, if the symmetry breaking is spontaneous,
then both the background $\bar k_{abc\cdots}$ and the background vierbein $\barvb a \mu$
arise dynamically as vacuum expectation values.
When the NG modes for both the broken diffeomorphism invariance
and local Lorentz invariance are included in the action,
the symmetry of $S$ under both sets of transformations is restored.

\section{Excitations}\label{sec3}

Effective gravitational field theories with background fields have
excitations that depend on the form of the symmetry breaking.

A theory with spontaneous diffeomorphism and local Lorentz violation
has excitations that occur as massless NG and massive Higgs excitations.
The question of whether a gravitational Higgs mechanism can occur
becomes relevant as well.

In contrast, in a theory with explicit diffeomorphism and local Lorentz breaking,
the background field is nondynamical and does not have excitations.
The fact that the background is not able to have backreactions and provides a
structure with  ``prior geometry'' is very different from GR and theories
with spontaneous spacetime symmetry breaking.
However, the loss of local symmetry in theories with explicit breaking
does give rise to additional degrees of freedom in the metric, 
and these additional excitations lead to modified gravitational interactions.

In theories with explicit breaking,
a St\"uckelberg approach is often used.
In this case, the multi-component background is replaced by a
function, including derivatives, of four dynamical scalar fields,
and the local spacetime symmetry is restored.
Excitations of the St\"uckelberg fields have the form of NG excitations.
Thus, the question of how these excitations 
compare with the NG excitations in a theory
with spontaneous breaking is pertinent.

\subsection{NG Modes}

Typically, a potential term $V (g_{\mu\nu}, k_{\la\mu\nu\cdots})$ in the Lagrangian induces
spontaneous spacetime symmetry breaking,
where the potential is formed from scalar combinations of $g_{\mu\nu}$ and $k_{\la\mu\nu\cdots}$
and possibly their derivatives and other fields as well.
The spontaneous diffeomorphism breaking occurs when a nonzero solution
$\bar k_{\la\mu\nu\cdots}$ and $\vev{g_{\mu\nu}}$ causes the potential 
to be at a minimum obeying $V^\prime =0$.
Massless NG modes then occur as excitations about the vacuum
solution that stay in the minimum (still obeying $V^\prime =0$), 
while massive Higgs excitations are solutions that
do not stay in the minimum (with $V^\prime \ne 0$).

The NG modes are generated by the broken symmetry transformations.
With broken diffeomorphims,
they have the form of infinitesimal Lie derivatives,
where the four parameters $\xi^\mu$ become the NG degrees of freedom.
The excitations can then be written at leading order as
\beq
k_{\la\mu\nu\cdots} \simeq \bar k_{\la\mu\nu\cdots} 
+ (D_\la \xi^\al) \bar k_{\al\mu\nu\cdots}
+ (D_\mu \xi^\al) \bar k_{\mu\al\nu\cdots}
+ \cdots
+ \xi^\al D_\al \bar k_{\al\mu\nu\cdots}
+ (\de k_{\la\mu\nu\cdots})_{\rm massive} ,
\label{diffNGmodes}
\eeq
where $(\de k_{\la\mu\nu\cdots})_{\rm massive}$ denotes the massive Higgs modes.
While there are only four NG modes associated with diffeomorphism breaking,
the number of massive modes depends on the type of tensor, the potential $V$,
and the kinetic terms for $k_{\la\mu\nu\cdots}$.
The question of whether ghost modes exist depends on these features as well.

To generate the NG modes for the broken local Lorentz transformations,
infinitesimal excitations having the form of broken Lorentz transformations
around the vacuum solution in a vierbein formalism can be used.
In this case, there are six NG excitations,
which can be written in terms of $\ep_{ab}$ as
\beq
k_{abc\cdots} \simeq \bar k_{abc\cdots} 
+ \ep_a^{\pt{a}j} \bar k_{jbc\cdots}
+ \ep_b^{\pt{a}j} \bar k_{ajc\cdots}
+ \cdots 
+ (\de k_{abc\cdots})_{\rm massive} ,
\label{lorentzNGmodes}
\eeq
where $ (\de k_{abc\cdots})_{\rm massive}$ are the components of
the massive excitations defined with respect to the local Lorentz frame.

In theories with spontaneous spacetime symmetry breaking,
the NG modes can be interpreted in some cases as known gauge fields,
such as photons or gravitons \cite{bjorken63,nambu68,bb1,bb2,cardinal}.
Alternatively,
the NG modes can be gauged into the vierbein,
which modifies the gravitational interactions.
If a Riemann--Cartan geometry is considered,
theories with a Higgs mechanism that gives rise to mass
terms for the spin connection become possible \cite{bb1,bb2}.
However, finding models that are free of ghosts remains elusive.

\subsection{St\"uckelberg Fields}

If a St\"uckelberg approach is used in a theory with explicit diffeomorphism breaking,
the nondynamical background is replaced by four dynamical scalars as shown in \rf{ktransf1},
which restores the local diffeomorphism invariance.
The St\"uckelberg version is dynamically equivalent to the original
explicit-breaking form, 
since imposing gauge-fixing conditions on the scalars, $\ph^A = \de^A_\mu x^\mu$,
reduces the expression on the right-hand side in \rf{ktransf1} back to $\bar k_{\la\mu\nu\cdots}$
and leaves the metric with four additional degrees of freedom.  

To restore diffeomorphism invariance, the excitations in the St\"uckelberg scalars
have the form of NG modes.
Writing the excitations as
\beq
\ph^A = \de^A_\mu (x^\mu + \xi^\mu) 
\label{stuckphi}
\eeq
and substituting them into the expression in \rf{ktransf1}
gives the leading order that
\beq
\partial_\la \ph^A \partial_\mu \ph^B 
\partial_\nu \ph^C \cdots \bar k_{ABC\cdots} (\ph) 
\simeq \bar k_{\la\mu\nu\cdots} 
+ (D_\la \xi^\al) \bar k_{\al\mu\nu\cdots}
+ (D_\mu \xi^\al) \bar k_{\mu\al\nu\cdots}
+ \cdots
+ \xi^\al D_\al \bar k_{\al\mu\nu\cdots} .
\label{Stuckexcit}
\eeq
This has the same form as the NG excitations about the background
$\bar k_{\la\mu\nu\cdots}$ in a corresponding theory with spontaneous breaking,
as given in \rf{diffNGmodes},
but where there are no massive excitations.  

An explicit-breaking theory in a St\"uckelberg description
can be viewed as a theory with spontaneous diffeomorphism breaking,
but where it is the scalar fields that acquire vacuum values
of the form $\ph^A = \de^A_\mu x^\mu$.
While these vacuum values replicate the original background $\bar k_{\la\mu\nu\cdots}$
as it appears in the Lagrangian,
there are still only four degrees of freedom in the explicit-breaking case.  
This is, in general, not sufficient to provide a vacuum solution for the multi-component 
background tensor $\bar k_{\la\mu\nu\cdots}$,
which must satisfy Euler--Lagrange equations for all of its components
if it is to be fully dynamical.
Thus, instead, \rf{nondyn} continues to hold for the explicit-breaking theory,
and $\bar k_{\la\mu\nu\cdots}$ remains nondynamical.

\section{Noether Identities}\label{sec4}

Field theories with local symmetries obey Noether's second theorem \cite{en18,Traut62},
which states that off-shell identities  relating the Euler--Lagrange expressions
for the dynamical fields must hold.
The~main consequence of the Noether identities is that
not all of the equations of motion are independent when there are local symmetries.

For example, in GR with matter fields $f^\ps$,
the Noether identities that result from diffeomorphism invariance have the form
\beq
D_\mu ( G^{\mu\nu} - T^{\mu\nu} ) 
+ \fr {\de {\cal L}} {\de f^\ps} \ga^{\ps\nu} + D_\mu (\fr {\de {\cal L}} {\de f^\ps} \ga^{\ps\mu\nu} ) = 0 ,
\label{NoetherGR}
\eeq
The coefficients $\ga^{\ps\nu}$ and $\ga^{\ps\mu\nu}$
denote functions of the field components,
where their specific form depends on the theory,
while $( G^{\mu\nu} - T^{\mu\nu} )$ and $ \fr {\de {\cal L}} {\de f^\ps}$ are
the Euler--Lagrange expressions for the metric and matter fields, respectively.
What this identity says is that four of the dynamical equations of motion are not independent
when there is local diffeomorphism invariance.
Alternatively, when this identity is combined with the contracted Bianchi identity,
which states that $D_\mu G^{\mu\nu} = 0$,
it shows that covariant energy-momentum conservation,
$D_\mu T^{\mu\nu}  = 0$, must automatically hold when the
dynamical matter fields are on-shell obeying $\fr {\de {\cal L}} {\de f^\ps} = 0$.

In a vierbein description, Noether identities resulting from local Lorentz invariance
hold as well.  
Using a vierbein description in GR with matter fields $f^\ps$,
the resulting Noether identities are
\beq
(G^{\mu\nu} - T^{\mu\nu}) (\lvb \mu a \lvb \nu b - \lvb \mu b \lvb \nu a)
+ \fr 1 2 \fr {\de {\cal L}} {\de f^y}  (X_{[ab]})_{\pt{y}x}^{y} f^x = 0 .
\label{LLINoether}
\eeq
In this case, since $G^{\mu\nu} = G^{\nu\mu}$ holds for the Einstein tensor,
the result of this identity is that the energy-momentum tensor in the vierbein
description must also be symmetric when the matter fields are on shell.

While effective gravitational theories with background fields break diffeomorphism
and local Lorentz invariance either explicitly or spontaneously,
it may seem that there are no Noether identities that apply in these theories.
However, if the theory is to remain observer independent,
the action must still be mathematically invariant under the observer
spacetime transformations.
Applying the observer spacetime transformations and imposing $\de S = 0$
therefore results in Noether identities that must hold even when there is a background field.
Under the observer transformations,
the background field transforms along with the metric and matter fields,
which can yield Noether identities for both observer diffeomorphisms and
local Lorentz transformations \cite{rbas16}.

For example, performing observer diffeomorphisms on the action in \rf{brokenSmetric}
and requiring that it be observer independent results in Noether identities of the form
\beq
D_\mu ( G^{\mu\nu} - T^{\mu\nu} ) 
+ \fr {\de {\cal L}} {\de f^\ps} \ga^{\ps\nu} + D_\mu (\fr {\de {\cal L}} {\de f^\ps} \ga^{\ps\mu\nu} ) 
+ \fr {\de {\cal L}} {\de \bar k_{\al\be\ga\cdots}} \la^{\nu}_{\al\be\ga\cdots} 
+ D_\mu (\fr {\de {\cal L}} {\de \bar k_{\al\be\ga\cdots}}  \la^{\mu\nu}_{\al\be\ga\cdots} ) 
= 0 .
\quad
\label{obsNoetherGR}
\eeq
In this case, $\ga^{\ps\nu}$, $\ga^{\ps\mu\nu}$, 
$\la^{\nu}_{\al\be\ga\cdots}$ and $ \la^{\mu\nu}_{\al\be\ga\cdots}$
all denote coefficients that are functions of the field components.
Notice in this case that Euler--Lagrange expressions for the background,
$\fr {\de {\cal L}} {\de \bar k_{\al\be\ga\cdots}}$, appear in this identity.
As a result of this,
the interpretation of the Noether identies when there is a background field present
depends on whether the symmetry breaking is explicit or spontaneous.

For the case of explicit breaking, 
the background $\bar k_{\la\mu\nu\cdots}$
is nondynamical and the Euler--Lagrange equation for it need not vanish,
as indicated in \rf{nondyn}.
Thus, when the metric and conventional matter fields are on shell,
theoretical consistency requires that the following equation must hold:
\beq
 \fr {\de {\cal L}} {\de \bar k_{\al\be\ga\cdots}} \la^{\nu}_{\al\be\ga\cdots} 
+ D_\mu (\fr {\de {\cal L}} {\de \bar k_{\al\be\ga\cdots}}  \la^{\mu\nu}_{\al\be\ga\cdots} ) 
= 0 .
\label{diffconstraint}
\eeq
This results in a different interpretation from GR,
since it is no longer an option to set the Euler--Lagrange expressions for
$\bar k_{\la\mu\nu\cdots}$ to zero.
Instead, it is the four additional metric modes that exist
as a result of the symmetry breaking that must satisfy this equation.
In some theories,
the couplings between the additional metric modes and the background
are insufficient to allow the conditions in~\rf{diffconstraint} to hold.
For example, if a particular ansatz form of the metric is chosen that does not
include any of the needed additional modes, then some backgrounds 
can become incompatible with the Noether~identities.

However, if the symmetry breaking is spontaneous,
then $\bar k_{\la\mu\nu\cdots}$ is a dynamical vacuum solution
and Euler--Lagrange equations hold, as in \rf{dynvac}.
If excitations are included as in \rf{diffNGmodes},
then the Euler--Lagrange equations for the tensor $k_{\la\mu\nu\cdots}$
(including the NG and massive excitations) hold.
Thus, with spontaneous diffeomorphism breaking,
the interpretation of the Noether identities is the same as in GR.
Four of the equations of motion are not dynamically independent,
and covariant energy-momentum conservation holds when
all of the dynamical fields are on shell.  

If a vierbein formalism is used,
the Noether identities resulting from observer local Lorentz invariance can be obtained.
In this case, the action is given in \rf{brokenSvier},
which depends on both the background field and a background vierbein.
Requiring that $S$ be unchanged under observer local Lorentz transformations gives
six Noether identities, which have the form
\begin{equation}
\begin{array}{cc}
(G^{\mu\nu} - T^{\mu\nu}) (\lvb \mu a \lvb \nu b - \lvb \mu b \lvb \nu a)
+ \fr 1 2 \fr {\de {\cal L}} {\de f^y}  (X_{[ab]})_{\pt{y}x}^{y} f^x 
+ \left( \fr {\de {\cal L}} {\de \barvb \mu a} \barlvb \mu b - \fr {\de {\cal L}} {\de \barvb \mu b} \barlvb \mu a \right) \\
+ \fr {\de {\cal L}} {\de \bar k_{cde\cdots}} \left[ (\et_{ac} \bar k_{bde\cdots} - \et_{bc} \bar k_{ade\cdots}) \right.
+ \left. (\et_{ad} \bar k_{cbe\cdots} - \et_{bd} \bar k_{cae\cdots}) \right. \\
+  \left. (\et_{ac} \bar k_{cdb\cdots} - \et_{bc} \bar k_{cda\cdots})
+ \cdots \right] = 0 .
\end{array}
\label{vierLLINoether}
\end{equation}
Again,
the interpretation of these identities depends on the type of symmetry breaking.

With explicit breaking of local Lorentz invariance,
the background field and the background vierbein are nondynamical and therefore
in general obey the relations
\beq
\fr {\de {\cal L}} {\de \barvb \mu a} \ne 0 ,
\quad\quad
\fr {\de {\cal L}} {\de \bar k_{abc\cdots}} \ne 0 .
\label{novanishvier}
\eeq
Thus, when the matter fields are on shell and the symmetry of the Einstein tensor is used,
it follows that the energy-momentum tensor in the vierbein description is symmetric only
if the remaining terms in \rf{vierLLINoether} combine to give zero.
With explicit Lorentz breaking, the vierbein has six additional degrees of freedom due
to the loss of the local symmetry.
It is these degrees of freedom that must make the remaining terms in \rf{vierLLINoether} vanish
in order for $T^{\mu\nu}$ to be symmetric.

On the other hand, if the breaking of local Lorentz invariance is spontaneous,
then both $\barvb \mu a$ and $\bar k_{abc\cdots}$ are dynamical vacuum solutions.
Therefore, the Euler--Lagrange equations for the vacuum hold,
and these equations continue to hold when excitations are included.
The result in this case is the same as in GR.
When the all the dynamical equations of motion hold,
$T^{\mu\nu}$ is automatically symmetric.

As described above,
the St\"uckelberg formalism allows an explicit-breaking theory
with a nondynamical background $\bar k_{\la\mu\nu\cdots}$
to be reinterpreted as a dynamical theory with four additional scalars $\ph^A$.
The St\"uckelberg approach is often referred to as a trick,
since it restores the symmetry in a theory where it is initially explicitly broken,
and it makes the theory dynamical.
Some insight into why this trick works can be obtained by examining the Noether identities
that hold with scalar fields.

If the St\"uckelberg substitution \rf{ktransf1} is made in the action in \rf{brokenSmetric},
the result is a new action that depends on the fields $g_{\mu\nu}$, $f^\ps$, and $\ph^A$.
The Noether identities stemming from diffeomorphism transformations in this case are
\beq
D_\mu ( G^{\mu\nu} - T^{\mu\nu} ) 
+ \fr {\de {\cal L}} {\de f^\ps} \ga^{\ps\nu} + D_\mu (\fr {\de {\cal L}} {\de f^\ps} \ga^{\ps\mu\nu} ) 
+ \left( -D_\mu \fr {\partial {\cal L}} {\partial \partial_\mu \ph^A}  
+ \fr {\partial {\cal L}} {\partial \ph^A}\right) \partial_\nu \ph^A = 0.
\label{phiA2}
\eeq
When the metric and matter fields are put on shell,
and assuming the derivatives $\partial_\nu \ph^A$ are linearly independent,
the resulting conditions that must hold are
\beq
-D_\mu \fr {\partial {\cal L}} {\partial \partial_\mu \ph^A}  
+ \fr {\partial {\cal L}} {\partial \ph^A}  = 0 .
\label{phiA3}
\eeq
These have the form of the Euler--Lagrange equations of motion for the scalars $\ph^A$.

It is important to realize that the Noether identities in \rf{phiA2} and the
conditions in \rf{phiA3} that follow from them can be obtained in two different ways.
In the first, the scalars $\ph^A$ are treated as fixed nondynamical background 
fields that explicitly break diffeomorphism invariance.
The substitution~\rf{ktransf1} in this case replaces the nondynamical background
$\bar k_{\la\mu\nu\cdots}$ by derivatives of nondynamical scalars $\ph^A$.
The Noether identities in \rf{phiA2} follow, in this case, from imposing the requirement of 
observer independence and using observer diffeomorphism transformations.
However, the same identities in \rf{phiA2} follow using the St\"uckelberg trick,
where the scalars $\ph^A$ in this case are treated as dynamical fields.
It is unbroken diffeomorphism invariance that gives rise to the Noether identities
in \rf{phiA2} in this~approach.

Thus, using either nondynamical or dynamical scalars $\ph^A$,
the result of the Noether identities is that the Euler--Lagrange equations in 
\rf{phiA3} must hold.
It is this fact that enables the St\"uckelberg approach to work.
Starting with a theory with explicit breaking and a nondynamical background,
the replacement in \rf{ktransf1} is made using fixed nondynamical scalars.
The consistency of the explicit-breaking theory requires that the conditions
stemming from the Noether identities must hold.
However, these have the form of the Euler-L-agrange equations for the scalars.
Letting the scalars be dynamical and restoring diffeomorphism invariance
is then possible because the Euler--Lagrange equations for the 
dynamical scalars are already imposed.

It is important to note as well that in the first case where the scalars $\ph^A$
are nondynamical,
the Euler--Lagrange equations in \rf{phiA3} must be satisfied by
the additional metric modes that occur due to the lose of local symmetry.
In contrast, in the St\"uckelberg approach, 
where the scalars are dynamical and the metric can be gauge fixed,
it is then the scalars themselves that must satisfy their own Euler--Lagrange equations,
as expected for fields that are dynamical.

\section{Summary and Conclusions}\label{sec5}

Effective gravitational field theories with background fields are used
in a variety of investigations looking at possible modifications of gravity,
quantum gravity effects, and phenomenological effects of spacetime symmetry breaking.
The presence of a background field breaks diffeomorphism and local Lorentz
invariance either explicitly or spontaneously,
and the behavior and interpretation of the background field depends
on which type of breaking occurs.

With explicit breaking, the background is nondynamical,
and it does not obey Euler--Lagrange equations of motion.
Noether identities in this case are obtained by imposing observer independence.
It is the additional metric modes that result from the absence of local symmetry
that must satisfy the Noether identities.

In contrast, with spontaneous breaking, the background is dynamical
and it obeys Euler--Lagrange equations either as a vacuum solution
or when the NG and massive Higgs modes are included.
Noether identities follow in this case from the unbroken local symmetry,
and their interpretation is similar to~GR.

A St\"uckelberg approach can be used to turn a theory with explicit
breaking and a nondynamical background into an equivalent theory
with four dynamical scalars where the symmetry is restored.
The St\"uckelberg excitations are NG modes about vacuum solutions
for the four scalars.  
The original fixed background $\bar k_{\la\mu\nu\cdots}$ remains
nondynamical and does not satisfy its Euler--Lagrange equations.
Instead, it is the scalar field Euler--Lagrange equations that hold.
The fact that these equations hold regardless of whether the scalars
are dynamical or nondynamical is a key feature that permits the
St\"uckelberg approach to work.

\vspace{6pt}



\begin{thebibliography}{999}

\bibitem{rb15a}
Bluhm, R.
Explicit versus spontaneous diffeomorphism breaking in gravity.
\emph{Phys. Rev. D} \textbf{2015}, \emph{91}, 065034.

\bibitem{sme1}
Colladay, D.; Kosteleck\'y, V.A.
CPT violation and the standard model.
\emph{Phys. Rev. D} \textbf{1997}, {\emph{ 55}}, 6760.

\bibitem{sme2}
Colladay,~D.; Kosteleck\'y, V.A.
Lorentz-violating extension of the standard model.
\emph{Phys. Rev. D} \textbf{1998}, {\emph{58}}, 116002.

\bibitem{sme3}
Kosteleck\'y,~V.A.; Lehnert, R.
Stability, causality, and Lorentz and CPT violation.
\emph{Phys. Rev. D} \textbf{2001}, {\emph{63}}, 065008. .

\bibitem{akgrav04}
Kosteleck\'y, V.A.
Gravity, Lorentz violation, and the standard model.
\emph{Phys. Rev. D} \textbf{2004}, {\emph{69}}, 105009.

\bibitem{rbsme}
Bluhm, R.
Overview of the SME: Implications and Phenomenology of Lorentz Violation.
In {\it Special Relativity: Will It Survive the Next 101 Years?}; Ehlers, J., L\"ammerzahl, C., Eds.;
Springer: Berlin, Germany, 2006.

\bibitem{ks1}
Kosteleck\'y, V.A.; Samuel, S.
Gravitational phenomenology in higher dimensional theories and strings.
\emph{Phys. Rev. D} \textbf{1989}, \emph{40}, 1886.

\bibitem{ks2}
Kosteleck\'y, V.A.; Samuel, S. Spontaneous breaking of Lorentz symmetry in string theory.
\emph{Phys. Rev. D} \textbf{1989}, \emph{39}, 683.

\bibitem{ks3}
Kosteleck\'y, V.A.; Samuel, S. Phenomenological Gravitational Constraints on Strings and Higher Dimensional Theories.
\emph{Phys. Rev. Lett.} \textbf{1989}, \emph{63}, 224. 

\bibitem{bb1}
Bluhm, R.; Kosteleck\'y, V.A.
Spontaneous Lorentz violation, Nambu-Goldstone modes, and gravity.
\emph{Phys. Rev. D}  \textbf{2005}, \emph{71}, 065008.

\bibitem{bb2}
Bluhm, R.; Fung, S.-H.; Kosteleck\'y, V.A. 
Spontaneous Lorentz and diffeomorphism violation, massive modes, and gravity.
\emph{Phys. Rev. D} \textbf{2008}, \emph{77}, 065020. 

\bibitem{rbngrpav} 
Bluhm, R.; Gagne, N.L.; Potting, R.; Vrublevskis, A.
Constraints and stability in vector theories with spontaneous Lorentz violation.
\emph{Phys. Rev. D}  \textbf{2008}, \emph{77}, 125007.

\bibitem{ms1}
Seifert, M.D.
Vector models of gravitational Lorentz symmetry breaking.
\emph{Phys. Rev. D} \textbf{2009} \emph{79}, 124012.

\bibitem{ms2}
Seifert,~M.D.
Generalized bumblebee models and Lorentz-violating electrodynamics.
\emph{Phys. Rev. D}  \textbf{2010}, \emph{81},~065010. 

\bibitem{ch14}
Hernaski, C.A.
Quantization and stability of bumblebee electrodynamics.
\emph{Phys. Rev. D}  \textbf{2014}, \emph{90}, 124036.

\bibitem{eu15}
Escobar, C.A.; Urrutia, L.F.
Photons emerging as Goldstone bosons from spontaneous Lorentz symmetry breaking: The Abelian Nambu model.
\emph{Phys. Rev. D}  \textbf{2015}, \emph{92}, 025042.

\bibitem{ee1}
Jacobson, T.; Mattingly, D.
Einstein-Aether Waves.
\emph{Phys. Rev. D}\textbf{2004},  \emph{70}, 024003.

\bibitem{ee2}
Jacobson, T.
Einstein-aether gravity: A status report.
\emph{arXiv} \textbf{2007}, arXiv:0801.1547. 

\bibitem{cardinal}
Kosteleck\'y, V.A.; Potting, R.
Gravity from local Lorentz violation.
\emph{Gen. Rel. Grav.}  \textbf{2005},  \emph{37}, 1675;
Gravity from spontaneous Lorentz violation.
Kosteleck\'y, V.A.; Potting, R. \emph{Phys. Rev. D}  \textbf{2009}, \emph{79}, 065018.

\bibitem{phon1}
Altschul, B.; Bailey, Q.G.; Kosteleck\'y, V.A.
Lorentz violation with an antisymmetric tensor.
\emph{Phys. Rev. D} \textbf{2010}, \emph{81}, 065028.

\bibitem{phon2}
Hernaski, C.
Spontaneous Breaking of Lorentz Symmetry with an antisymmetric tensor.
\emph{Phys. Rev. D}  \textbf{2016}, \emph{94}, 105004. 

\bibitem{rjsp1}
Jackiw, R.; Pi, S.-Y.
Chern-Simons modification of general relativity.
\emph{Phys. Rev. D} \textbf{2003}, {68}, 104012.

\bibitem{rjsp2}
Jackiw, R.
Lorentz violation in a diffeomorphism-invariant theory.
In {\it CPT and Lorentz Symmetry IV};   Kosteleck\'y, V.A., Ed.;
World Scientific: Singapore, 2008. 

\bibitem{mg1}
Hinterbichler, K.
Theoretical Aspects of Massive Gravity.
\emph{Rev. Mod. Phys.} \textbf{2012}, \emph{84}, 671.

\bibitem{mg2}
 De Rham, C.
Massive Gravity.
\emph{Living Rev. Relativ.} \textbf{2014}, \emph{17}, 7. 

\bibitem{varcoups1}
Kosteleck\'y, V.A.; Lehnert, R.; Perry, M.J.
Spacetime-varying couplings and Lorentz violation.
\emph{Phys. Rev. D}  \textbf{2003}, {68}, 123511.

\bibitem{varcoups2}
Bertolami, O.; Lehnert, R.; Potting, R.; Ribeiro, A.
Cosmological acceleration, varying couplings, and Lorentz breaking.
\emph{Phys. Rev. D} \textbf{2004}, { 69}, 083513. 


\bibitem{tvar1}
Ginges, J.S.M.; Flambaum, V.V.
Violations of fundamental symmetries in atoms and tests of unification theories of elementary particles.
\emph{Phys. Rept.} \textbf{2004}, \emph{397}, 63.

\bibitem{tvar2}
Uzan, J.-P.
Varying Constants, Gravitation and Cosmology.
\emph{Living Rev. Relativ.} \textbf{2011}, \emph{14}, 2.

\bibitem{tvar3}
Sol\`a, J.
Fundamental Constants in Physics and Their Time Variation.
\emph{Mod. Phys. Lett. A}  \textbf{2015}, \emph{30}, 1502004.  

\bibitem{rb15b}
Bluhm, R.
Spacetime symmetry breaking and Einstein-Maxwell theory.
\emph{Phys. Rev. D} \textbf{2015}, \emph{92}, 085015.

\bibitem{rbas16}
Bluhm, R.; \v Sehi\'c, A.
Noether identities in gravity theories with nondynamical backgrounds 
and explicit spacetime symmetry breaking.
\emph{Phys. Rev. D} \textbf{2016}, \emph{94}, 104034.

\bibitem{CPTmeetings}
For reviews of experimental and theoretical
approaches to violations of fundamental spacetime symmetries.
In {\it CPT and Lorentz Symmetry VII};  Kosteleck\'y, V.A., Ed.;
World Scientific: Singapore, 2016.

\bibitem{m05}
Mattingly, D.
Modern tests of Lorentz invariance. 
\emph{Living Rev. Relativ.} \textbf{2005}, \emph{8}, 5.

\bibitem{ML12}
Mattingly, D.; Liberati, S.
Lorentz breaking effective field theory models for matter and gravity: Theory and observational constraints. \emph{arXiv} \textbf{2012}, 
arXiv:1208.1071. 

\bibitem{JT14}
Tasson, J.
What do We Know about Lorentz Invariance?
\emph{Rept. Prog. Phys.} \textbf{2014}, \emph{77}, 062901.

\bibitem{RB14}
Bluhm, R. 
Observable Constraints on Local Lorentz Invariance.
In {\it Springer Handbook of Spacetime};  Ashtekar,~A.,  Petkov, V., Eds.;
Springer: Berlin, Germany, 2014.

\bibitem{Iorio1}
Iorio, L. 
Editorial for the Special Issue 100 Years of Chronogeometrodynamics: The Status of the Einstein's 
Theory of Gravitation in Its Centennial Year.
\emph{Universe} \textbf{2015}, \emph{1}, 38.

\bibitem{will}
Will, C.M. 
The Confrontation between General Relativity and Experiment.
\emph{Living Rev. Relativ.} \textbf{2014}, \emph{17}, 4.

\bibitem{hees}
Hees, A.; Bailey, Q.G.; Bourgoin, A.; Bars, H.P.; Guerlin, C.; le Poncin-Lafitte, C.; 
Tests of Lorentz symmetry in the gravitational sector.
\emph{Universe} \textbf{2016}, \emph{2}, 30.

\bibitem{smoot} 
Debono, I.; Smoot, G.F. 
General Relativity and Cosmology: Unsolved Questions and Future Directions. 
\emph{Universe} \textbf{2016}, \emph{2}, 23.

\bibitem{stubbs}
Battat, J.B.R.; Chandler, J.F.; Stubbs, C.W.
Testing for Lorentz Violation: Constraints on Standard-Model Extension Parameters via Lunar Laser Ranging.
\emph{Phys. Rev. Lett.} \textbf{2007},  \emph{99}, 241103.

\bibitem{mueller1}
Mueller, H.; Chiow, S.-W.; Herrmann, S.; Chu, S.; Chung, K.-Y.
Atom Interferometry tests of the isotropy of post-Newtonian gravity.
\emph{Phys. Rev. Lett.} \textbf{2008}, \emph{100}, 031101.

\bibitem{mueller2}
Chung, K.-Y.; Chiow, S.-W.; Herrmann, S.; Chu, S.; Mueller, H.
Atom interferometry tests of local Lorentz invariance in gravity and electrodynamics.
\emph{Phys. Rev D} \textbf{2009}, \emph{80}, 016002.   

\bibitem{iorio2}
Iorio, L.
Orbital effects of Lorentz-violating Standard Model Extension gravitomagnetism around a static body: A sensitivity analysis.
\emph{Class. Quant. Gravit.} \textbf{2012}, \emph{29}, 175007.

\bibitem{yagi}
Yagi, K.; Blas, D.; Yunes, N.; Barausse, E.
Strong Binary Pulsar Constraints on Lorentz Violation in Gravity.
\emph{Phys. Rev. Lett.} \textbf{2014}, \emph{112}, 161101.

\bibitem{shao}
Shao, L.
Tests of local Lorentz invariance violation of gravity in the standard model extension with pulsars.
\emph{Phys. Rev. Lett.} \textbf{2014}, \emph{112}, 111103;
Shao, L. New pulsar limit on local Lorentz invariance violation of gravity in the standard-model extension.
\emph{Phys. Rev D} \textbf{2014}, \emph{90}, 122009.

\bibitem{angonin1}
Bourgoin, A.; Hees, A.; Bouquillon, S.; le Poncin-Lafitte, C.; Francou, G.; Angonin, M.-C.
Testing Lorentz symmetry with Lunar Laser Ranging.
\emph{Phys. Rev. Lett.} \textbf{2016}, \emph{117}, 241301.

\bibitem{angonin2}
Bourgoin, A.; le Poncin-Lafitte, C.; Hees, A.; Bouquillon, S.; Francou, G.; Angonin, M.-C.
Lorentz symmetry violations from matter-gravity couplings with Lunar Laser Ranging.
\emph{arXiv} \textbf{2017}, arXiv:1706.06294.  

\bibitem{aknr-tables}
Kosteleck\'y, V.A.; Russell, N.
Data tables for Lorentz and CPT violation.
\emph{Rev. Mod. Phys.} \textbf{2011}, \emph{83}, 11.

\bibitem{qbak06}
Bailey, Q.G.; Kosteleck\'y, V.A.
Signals for Lorentz violation in post-Newtonian gravity.
\emph{Phys. Rev. D} \textbf{2006}, \emph{74},~045001.

\bibitem{akjt1}
Kosteleck\'y, V.A.; Tasson, J.D.
Prospects for Large Relativity Violations in Matter-Gravity Couplings.
\emph{Phys. Rev. Lett.} \textbf{2009}, \emph{102}, 010402.

\bibitem{akjt2}
Kosteleck\'y, V.A.; Tasson, J.D. Matter-gravity couplings and Lorentz violation.
\emph{Phys. Rev. D}  \textbf{2011}, \emph{83}, 016013.

\bibitem{akjt3}
Kosteleck\'y, V.A.; Tasson, J.D. Constraints on Lorentz violation from gravitational Cherenkov radiation.
\emph{Phys. Lett. B} \textbf{2015}, \emph{749}, 551. 

\bibitem{hb14}
Hernaski, C.A.; Belich, H.
Lorentz violation and higher-derivative gravity.
\emph{Phys. Rev. D} \textbf{2014}, \emph{89}, 104027.

\bibitem{qbakrx}
Bailey, Q.G.; Kosteleck\'y, V.A.; Xu, R.
Short-range gravity and Lorentz violation.
\emph{Phys. Rev. D} \textbf{2015}, \emph{91},~022006.

\bibitem{cl16}
Lane, C.D. 
Spacetime variation of Lorentz-violation coefficients at a nonrelativistic scale.
\emph{Phys. Rev. D} \textbf{2016}, \emph{94}, 025016.

\bibitem{combined}
Shao, C.-G.; Tan, Y.-J.; Tan, W.-H.; Yang, S.-Q.; Luo, J.; Tobar, M.E.; Bailey, Q.G.; Long, J.C.; Weisman, E.; Xu,~R.; et al.
Combined search for Lorentz violation in short-range gravity.
\emph{Phys. Rev. Lett.} \textbf{2016}, \emph{117}, 071102.

\bibitem{ksmm1}
Kosteleck\'y, V.A.; Mewes, M.
Testing local Lorentz invariance with gravitational waves.
\emph{Phys. Lett. B} \textbf{2016}, \emph{757}, 510.

\bibitem{ksmm2}
Testing local Lorentz invariance with short-range gravity.
\emph{Phys. Lett. B} \textbf{2017}, \emph{766}, 137.

\bibitem{ksmm3}
Kosteleck\'y,~V.A.; Melissinos, A.C.; Mewes, M.
Searching for photon-sector Lorentz violation using gravitational-wave detectors
\emph{Phys. Lett. B} \textbf{2016}, \emph{761}, 1. 

\bibitem{ags03}
Arkani-Hamed, N.; Georgi, H.; Schwartz, M.D.
Effective field theory for massive gravitons and gravity in theory space. 
\emph{Ann. Phys.} \textbf{2003}, {\emph{305}}, 96.

\bibitem{bjorken63}
Bjorken, J.D.
A Dynamical origin for the electromagnetic field.
\emph{Ann. Phys.} \textbf{1963}, \emph{24}, 174.

\bibitem{nambu68}
Nambu, Y.
Quantum electrodynamics in nonlinear gauge.
\emph{Prog. Theor. Phys. Suppl. E} \textbf{1968}, \emph{68}, 190. 

\bibitem{en18}
Noether, E.
Invariante Variationsprobleme (Invariant Variation Problems). 
\emph{Nachr. D. K\"onig. Gesellsch. D. Wiss.
G\"ottingen: Math-Phys. Klasse} \textbf{1918}, \emph{II}, 235--257. (In German) 

\bibitem{Traut62} 
Trautman, A.
Conservation laws in general relativity. 
In {\it Gravitation: An Introduction to Current Research}; Witten, L., Ed.;
J. Wiley: New York, NY, USA, 1962.



\end{thebibliography}
\end{document}